\documentclass[12pt,times,authoryear]{elsarticle}

\usepackage[main=english,portuguese]{babel}
\usepackage[utf8]{inputenc}
\usepackage{amssymb}
\usepackage{amsmath}
\usepackage{siunitx}
\usepackage[final]{changes}
\colorlet{Changes@Color}{red}

\journal{Chemosphere}

\begin{document}

\begin{frontmatter}

\title{Oxidation of clofibric acid in aqueous solution using a non-thermal plasma discharge or gamma radiation}

\author[C2TN]{Joana Madureira}
\author[DSC]{Elisa Ceriani}
\author[IPFN]{Nuno Pinhão\corref{cor1}}\ead{npinhao@ctn.tecnico.ulisboa.pt}
\author[DSC]{Ester Marotta\corref{cor1}}\ead{ester.marotta@unipd.it}
\author[C2TN]{Rita Melo}
\author[C2TN]{Sandra Cabo Verde}
\author[DSC]{Cristina Paradisi}
\author[C2TN]{Fernanda M.A. Margaça}

\address[C2TN]{Centro de Ciências e Tecnologias Nucleares, Instituto Superior Técnico, Universidade de Lisboa, E.N. 10 ao km 139.7, 2695-066 Bobadela LRS, Portugal}
\address[DSC]{Dipartimento Scienze Chimiche, Università degli Studi di Padova, Via Marzolo 1, 35131 Padova, Italy}
\address[IPFN]{Instituto de Plasmas e Fusão Nuclear, Instituto Superior Técnico, Universidade de Lisboa, Av. Rovisco Pais, 1049-001 Lisboa, Portugal}

\cortext[cor1]{Corresponding authors}

\selectlanguage{english}

\begin{abstract}
In this work, we study degradation of clofibric acid (CFA) in aqueous solution using either ionizing radiation from a $^{60}$Co source or a non-thermal plasma produced by discharges in the air above the solution. The results obtained with the two technologies are compared in terms of effectiveness of CFA degradation and its by-products. In both cases the CFA degradation follows a quasi-exponential decay in time well modelled by a kinetic scheme which considers the competition between CFA and all reaction intermediates for the reactive species generated in solution as well as the amount of the end product formed. A new degradation law is deduced to explain the results. Although the end-product CO$_2$ was detected and the CFA conversion found to be very high under the studied conditions, HPLC analysis reveals several degradation intermediates still bearing the aromatic ring with the chlorine substituent. The extent of mineralization is rather limited. The energy yield is found to be higher in the gamma radiation experiments.
\end{abstract}


\end{frontmatter}

\section*{Highlights}
\begin{itemize}
\item The degradation of clofibric acid (CFA) in aqueous solution was studied.
\item Gamma radiation and a non-thermal plasma were used to promote advanced oxidation processes.
\item A new law is proposed to explain the degradation kinetics.
\item The degradation products of CFA were assessed.
\item The energy was significantly higher for the gamma radiation set-up than for plasma.
\end{itemize}


\section{Introduction}
\label{sec:I}

Pharmaceutical products are important emerging pollutants due to their large variety and increasing consumption in recent years. Due to the inadequate efficiency of conventional treatment processes, some are released into the aquatic environment as primary pollutants, metabolites and transformation products formed during wastewater treatments~\added{\citep{Patrolecco2015,Santos20091509,Salgado2862}} leading to contamination of surface waters, seawaters, groundwater and in some cases also drinking waters~\added{\citep{Salgado2012,Pereira2015108,deJesusGaffney2015199,Lolic2015240}}. They can have toxic effects~\added{\citep{Santos201045}} on humans after consumption of contaminated water or of food irrigated with polluted water, and affect aquatic organisms and other animal species. 

Clofibric acid (CFA) is the metabolite of the lipid regulators clofibrate, etofibrate and etofyllinclofibrate, used to decrease the level of cholesterol and triglycerides. It was detected in wastewater treatment plants~\added{\citep{Salgado2862,Dordio20091156}} in concentrations of up to \SI{41.4}{\micro\gram/L}~\added{\citep{Salgado20112359}}, representing one of the most abundant pharmaceutical compounds in the environment. CFA, which is expected to be fully ionized in natural waters (estimated pKa is 3.18), is non-biodegradable, highly mobile and has high persistence in the environment, with a half-life of \SI{21}{years}~\added{\citep{Winkler20013197}}.

Advanced oxidation processes (AOPs) use the hydroxyl radical ($\cdot$OH) for oxidation and are employed in novel technologies for water treatment and especially for the elimination of pharmaceutical products. Different AOPs have been used for this purpose, such as ozonation~\added{\citep{Huber03}}, Fenton and photo-Fenton processes~\added{\citep{Komtchou2015}}, photocatalysis~\added{\citep{Elmolla201046}} and ionizing radiations such as gamma radiation and electron-beam and, more recently, non-thermal plasmas (NTPs) produced by electrical discharges inside or in contact with the water to be treated~\added{\citep{Magureanu2015124}}. For example,~\added{\citet{Mezyk08}} developed a kinetic model to rationalize the effects of gamma radiation on three types of contaminants, particularly pharmaceuticals, and proposed the degradation mechanisms. \added{\citet{Illes20121479}} observed an effective degradation of ketoprofen by gamma radiation with no toxicity with \SI{5}{kGy} dose. The decomposition of diclofenac was studied by gamma radiation~\added{\citep{Liu2011}} using different initial conditions and by pulsed corona discharge~\added{\citep{Dobrin2013389}} and its degradation pathway was proposed. \added{\citet{Magureanu20113407}} studied the decomposition of three antibiotics (amoxicillin, oxacillin and ampicillin) using a dielectric barrier discharge (DBD) reaching their complete degradation after \SI{10}{\minute} for amoxicillin and \SI{30}{min} for oxacillin and ampicillin, respectively.
In recent years, CFA degradation has been studied using different AOPs. Ultraviolet (UV), vacuum ultraviolet, UV/TiO$_2$~\added{\citep{Li2012}}, ozonation~\added{\citep{Rosal20091061}}, different Fenton conditions~\added{\citep{Sires2007373}}, ionizing radiation such as electron beam~\added{\citep{Razavi09}} or gamma rays~\added{\citep{Csay201472}} and plasma from corona discharge~\added{\citep{Krause2011333}} were reported to be efficient in the elimination of CFA. Mechanisms for CFA decomposition have been proposed. However, the intermediates formed by plasma treatment were not clarified.

In this work, we report and discuss the results of studies of the oxidation of clofibric acid using two technologies: gamma radiation using a $^{60}$Co source and a non-thermal plasma (NTP) in contact with the liquid. We discuss the decomposition rates, the decomposition products and the energy yield for the two technologies.

\section{Material and Methods}
\label{sec:MM}

\subsection{Chemicals}

Clofibric acid, 4-chlorophenol and 2-hydroxyisobutyric acid, formic acid and acetonitrile were purchased from \added{Fluka}. Ultrapure grade water (milli-Q water) was obtained by filtration of deionized water with a \added{Millipore} system. Synthetic air (\SI{80}{\percent} nitrogen and \SI{20}{\percent} oxygen), used in the plasma experiments, was obtained from \added{Air Liquide}, with specified impurities of H$_2$O (\SI{<3}{ppm}) and of C$_n$H$_m$ (\SI{<0.5}{ppm}).
\begin{figure}[ht]
\begin{center}
\includegraphics[width=0.4\textwidth]{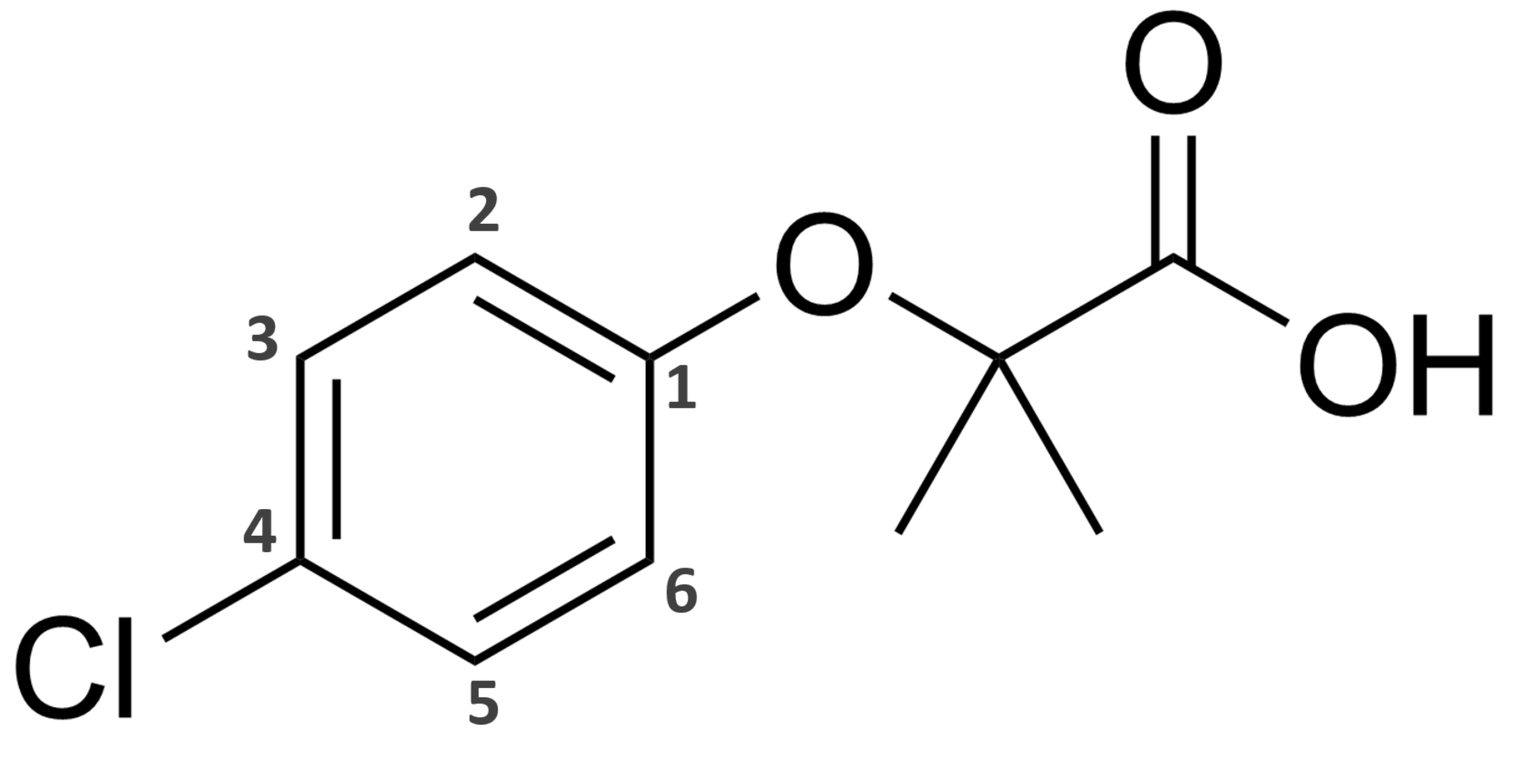}
\label{fig:CFA}
\caption{Structure of the investigated compound, clofibric acid.}
\end{center}
\end{figure}

\subsection{Preparation of clofibric acid solutions}

Solutions of desired concentration of \SIlist[fixed-exponent=-3,list-units=brackets,scientific-notation=fixed]{1e-4;5e-4;1e-3}{M} were prepared by dissolving CFA in Milli-Q water, without pH adjustment.

\subsection{Experiments of advanced oxidation of clofibric acid}

\subsubsection{Gamma radiation treatment}

Irradiation experiments of clofibric acid solutions were carried out in an equipment previously described by~\added{\citet{Madureira2017139}}. The irradiation chamber was a cavity of stainless steel (\SI{50}{cm} depth, \SI{20}{cm} width and \SI{65}{cm} height) with four $^{60}$Co sources placed in four channels on the side walls of the chamber. The irradiations were performed at room temperature with a dose rate of \SI{1.8}{kGy/h} and different absorbed doses (from \SIrange{0.1}{10}{kGy}). The samples were irradiated in vials with \SI{3}{mL} of solution in an
automatic rotation system to guarantee dose uniformity. The absorbed dose was measured by routine dosimeters~\added{\citep{Whittaker2001101}} while the local dose rate had been previously determined by the Fricke method~\added{\citep{ASTM:E1026}}. The treated solutions were analysed as described in the next paragraphs.

\subsubsection{Non-thermal plasma (NTP) treatment}

The experimental apparatus was previously described in detail by~\added{\citet{Marotta:PPAP201100036}}. Briefly, it includes the plasma reactor, the power supply, the gas line used to maintain a constant flow (\SI{30}{mL/min}) of humidified air above the treated solution and the instrumentation for electrical and chemical diagnostics. The reactor has a glass base (\SI{95x75x60}{mm}) closed by a teflon cap holding the active electrode (two parallel stainless steel wires with $\phi = \SI{0.15}{mm}$)  suspended \SI{10}{mm} above the solution to be treated (\SI{70}{mL}). The cap is equipped with openings for air inlet and outlet and for the withdrawal of aliquots (\SI{0.5}{mL}) of the solution at desired treatment times for off-line analyses, as described in the next paragraphs. Non-thermal plasma was generated by an ac voltage (\SI{18}{kV}, \SI{50}{Hz}) applied between the active electrode and the grounded electrode, which is placed in contact with the external surface of the glass bottom of the reactor. During the experiments, the peak voltage was kept constant and voltage and current profiles were monitored to assure the reproducibility of the electrical conditions. The gas exiting the reactor was subjected to online FT-IR analysis (Nicolet 5700) using a \SI{10}{cm} long flow cell with CaF$_2$ windows~\added{\citep{Marotta:EPJAP2011}}.

\subsection{Chemical analyses}

\subsubsection{Degradation studies}

Samples of solutions treated by both technologies were analysed by HPLC (\added{Shimadzu} LC-10AT pump with a UV–Vis \added{Shimadzu} SPD-10 detector) using a Merck Lichrospher 100 RP-18 (\SI{5}{\micro m}, \SI{250}{mm} $\times$ \SI{4.0}{mm}) column with the detection done at \SI{278}{nm}. The mobile phase used was A: \SI{0.1}{\percent} aqueous formic acid, and B: acetonitrile solution with \SI{5}{\percent} of water and \SI{0.1}{\percent} of formic acid. The flow rate was \SI{1}{mL/min}, the column temperature was maintained at \SI{298}{K} and the injection volume was \SI{20}{\micro L}. The gradient elution initial conditions were \SI{10}{\percent} B with linear gradient to \SI{40}{\percent} in \SI{20}{min}, being maintained for \SI{10}{min} (\SIrange[range-phrase=--,range-units=single]{20}{30}{min}), then followed by a return to the initial conditions within \SI{5}{min} (\SIrange[range-phrase=--,range-units=single]{30}{35}{min}) and kept \SI{5}{min} (\SIrange[range-phrase=--,range-units=single]{35}{40}{min}) for the chromatograph column equilibrium.

\subsubsection{Identification of decomposition products}

For the identification of the decomposition products, selected samples were analysed using an HPLC system (\added{Agilent Technologies} 1100 series) connected to a diode array and a mass spectrometer detector (\added{MSD} SL Trap). The eluent mixture was the same as used in the degradation studies. The ionization was performed within an electrospray (ESI) source alternating positive and negative polarity with the following parameters: nebulizer \SI{50}{psi}, dry gas flow rate \SI{8}{L/min}, dry gas temperature \SI{350}{\degree C}. The assignment of the intermediate products was confirmed by comparing their retention time, UV–Vis and mass spectra with those obtained from standard compounds, when available. Otherwise, peaks were tentatively identified comparing the obtained information with available data reported in the literature.

\section{Results}
\label{sec:R}

\subsection{Decomposition Rates}
\label{sec:3.1}

\begin{figure}[ht]
\centering\includegraphics[width=0.65\textwidth]{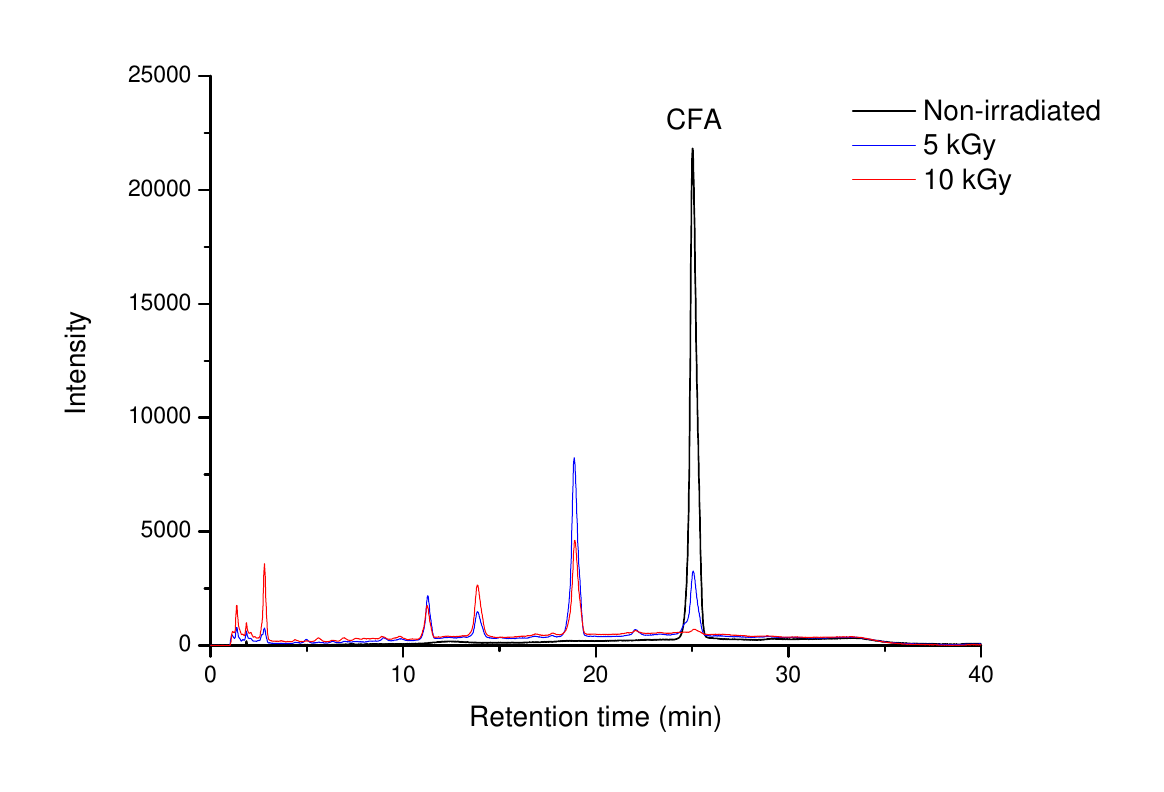}
\caption{ Chromatograms of solutions, with an initial clofibric acid concentration of \SI{1e-3}{M}, irradiated on a \added{$^{60}$Co} gamma source, as a function of the absorbed dose. The dose rate in the samples was \SI{1.8}{kGy/h}. }
\label{fig:F2}
\end{figure}

Figure \ref{fig:F2} shows the chromatograms of a CFA solution with an initial concentration of \SI{1e-3}{M}, before and after treatment with different doses of gamma radiation.
The peak area corresponding to CFA shows a considerable decrease with the absorbed dose indicating that degradation of the compound was taking place. At the same time, other peaks, attributed to decomposition products, appear and change in intensity with the absorbed doses. Similar observations were made on the treatment of CFA in the NTP reactor. 

\begin{figure}[ht]
\begin{center}
\includegraphics[height=0.58\textheight]{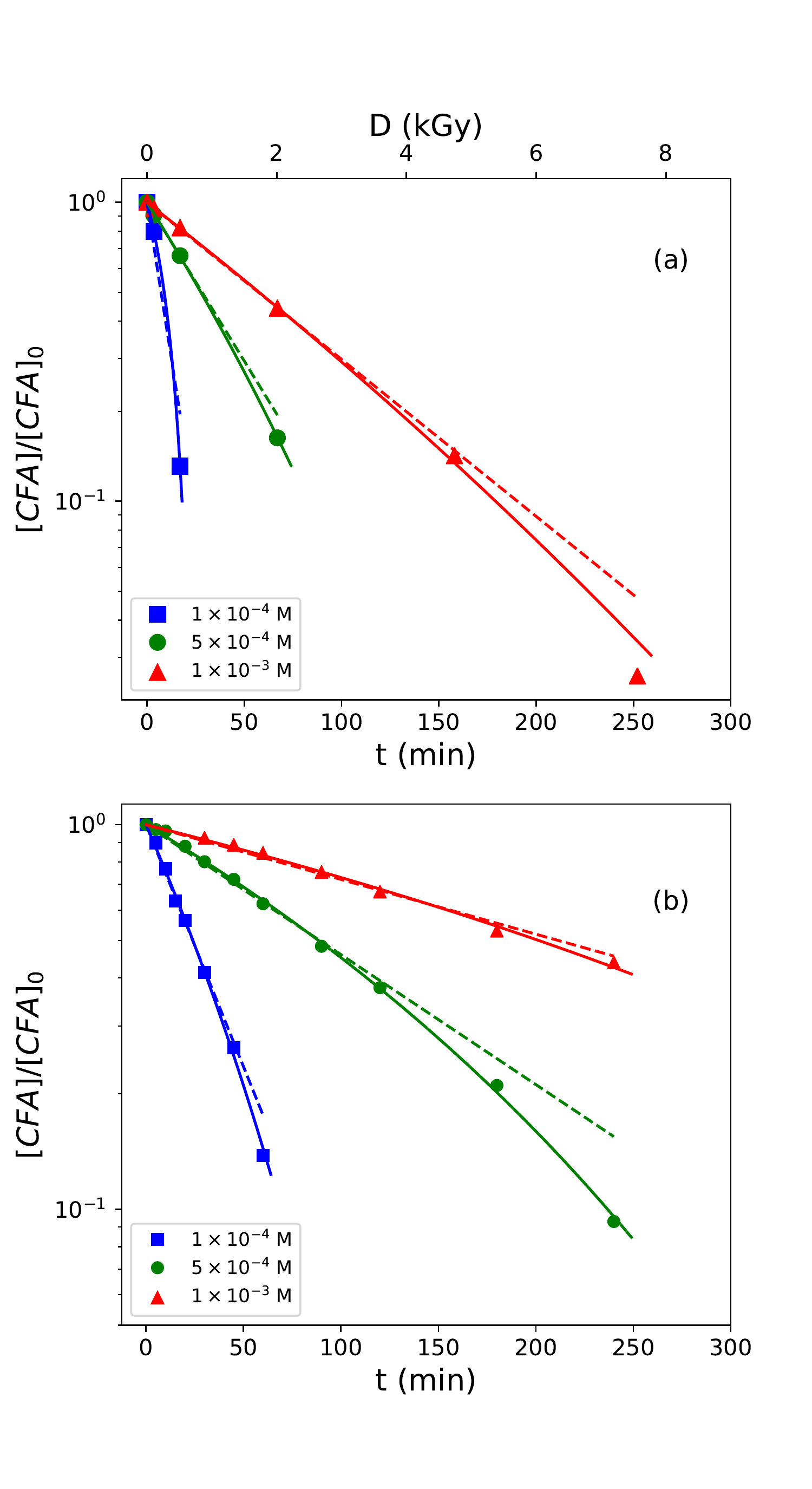}
\caption{Degradation of clofibric acid at different initial concentrations using: (a) gamma irradiation and (b) non-thermal plasma treatment. Experimental values (points), exponential fit (dotted lines) and fit to the new kinetic model described in  section~\ref{sec:4.1} (solid lines).}
\label{fig:F3}
\end{center}
\end{figure}

Figure~\ref{fig:F3} shows the time evolution of the ratio between the residual and initial concentrations of clofibric acid ([CFA]/[CFA]$_0$), treated by both technologies, for the different concentrations studied.

Under the present conditions, the degradation with gamma radiation was faster than with plasma. With gamma radiation (Figure~\ref{fig:F3}-a), after approximately \SI{60}{min} (corresponding to an absorbed dose of \SI{2}{kGy}) the degradation was \SIlist[list-units=brackets]{~100;84;56}{\percent}, for CFA solutions with initial concentrations of \SIlist[list-units=brackets,fixed-exponent=-3,,scientific-notation=fixed]{1e-4;5e-4;1e-3}{M}, respectively.\footnote{In the case of an initial concentration of \SI{1e-4}{M}, the final CFA concentration was below the detection limit.} For the non-thermal plasma at the same treatment time, the corresponding values were \SIlist[list-units=brackets]{~86;38;16}{\percent} (Figure 3-b)).
\added{\citet{Csay201472}} observed the same behaviour for the degradation of \SI{1e-4}{M} CFA with gamma radiation and above \SI{0.7}{kGy} (dose rate: \SI{12}{kGy/h}) no more CFA was detected. \added{\citet{Krause2011333}} reported the complete decomposition of CFA (with an initial concentration of \SI{1e-4}{M}) in \SI{30}{\minute} at \SI{500}{W} using a corona discharge with two barrier electrodes.

The experimental results were initially fit to an exponential law,
\begin{equation}
\frac{[\mathrm{CFA}]}{[\mathrm{CFA}]_0}=\exp(-\lambda t)\,,
\label{eq:Lambda}
\end{equation}
where $\lambda$ is the degradation coefficient for the process. 
The dotted lines in Figure~\ref{fig:F3} are the fit to equation (\ref{eq:Lambda}), neglecting the last points in each data set. As Figures~\ref{fig:F3}-a and \ref{fig:F3}-b show, the experimental data is well described by an exponential law, in particular for points with [CFA]/[CFA]$_0$ ratios above \num{0.2}. As the ratio decreases, however, the points deviate from the exponential fitting curve.

With both techniques, the rate of clofibric acid decomposition depends on its initial concentration in solution, being highest for the lowest concentration and decreasing with increasing initial concentration. 
The same type of dependency has also been found in studies of different compounds using electron-beam~\added{\citep{Slater:JAP1981}} or non-thermal plasmas~\added{\citep{Magureanu20103445, Marotta:PPAP201100036}}. \added{\citet{Slater:JAP1981}} have proposed a model to explain these observations leading to a dependency of the degradation constant with the reciprocal of the initial concentration of the compound.

\subsection{Decomposition products}\label{sec:DecPr}

The CFA solutions, treated by either gamma irradiation or non-thermal plasma, were subjected to HPLC-UV,  HPLC-MS and MS/MS analyses. A typical set of HPLC-UV traces recorded during a gamma radiolysis experiment is shown in Figure~\ref{fig:F2}. CFA and all of its products give strong mass spectrometric signals under negative ionization conditions whereas only some of them are detected under positive ionization conditions. Thus, the measurements were done mainly in negative ionization mode under which conditions CFA generates a peak at $m/z$~213 corresponding to the carboxylate anion (i.e. the acid conjugate base).

The chromatograms from samples produced by gamma radiation or non-thermal plasma treatment show  many similarities but also some differences in the detected products. 
Whenever possible, the products were identified with the use of standards. When standards were not available, MS and MS/MS spectra of the unknown products (the later spectra containing structural information derived from the detected fragments) were analysed and compared with those of CFA in search of informative similarities and differences. 

\begin{figure}[ht]
\begin{center}
\includegraphics[width=0.50\textwidth]{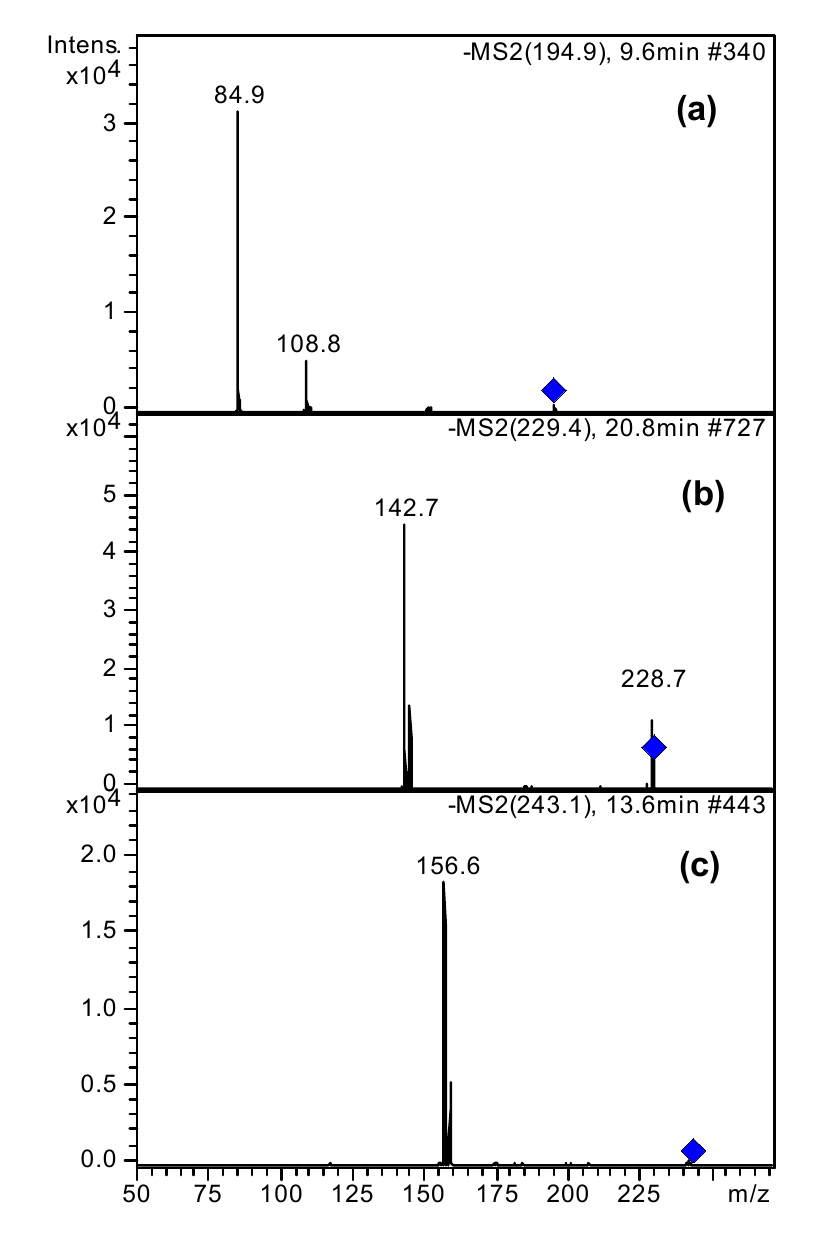}
\caption{Negative ion MS/MS spectra of intermediates (a) B , (b) E  and (c) G detected in the treatment of \SI{1e-3}{M} CFA with gamma radiation and/or with air non-thermal plasma (see Table~\ref{tbl:LCMS-P} and Scheme~\ref{fig:Prod-G} for details).}
\label{fig:MS_MS}
\end{center}
\end{figure}

For gamma treatment the major peaks are X (m/z 235), which remains unknown, B (m/z 195), C (m/z 143), and D (m/z 127). D was identified as 4-chlorophenol by matching data obtained in the analysis of an authentic standard. B was tentatively identified as 2-(4-hydroxyphenoxy)-isobutyric acid, the product of CFA hydroxydechlorination, based also on the MS/MS spectrum, where hydroxyphenolate (m/z 109) and methacrylate (m/z 85) are observed as ionic fragments (Figure~\ref{fig:MS_MS}a). C is attributed to 4-chlorocatechol, possibly formed by hydroxylation of 4-chlorophenol (D). In addition, two minor peaks, A and E, were also detected and characterized. A (m/z 103) was identified as 2-hydroxyisobutyric acid by matching data obtained in the analysis of an authentic standard. E (m/z 229) shows a fragment with m/z 143 in the MS/MS spectrum which suggests that this should be a product of CFA ring hydroxylation (Figure~\ref{fig:MS_MS}b). 4-chlorocatechol (m/z 143, C) could form from either D or E via reaction with hydroxyl radicals. For plasma treatment the UV chromatogram shows seven major peaks which correspond to CFA and to six degradation products (A, Z, F, G, D and E). Besides the common products also observed for gamma treatment (A, D and E), two additional products were detected, F and G with m/z 157 and m/z 243, respectively. As the molecular ion of CFA and that of product G differs by 30 a.m.u., this product is possibly a quinonic derivative of CFA. The appearance of a fragment ion with m/z 157 in the MS/MS spectrum of the ion at m/z 243 (G), corresponding to the loss of H$_2$C$=$C(CH$_3$)COOH, supports this hypothesis (Figure~\ref{fig:MS_MS}c). A subsequent reaction with OH radicals leads to the formation of F.
Most of the above products were also reported by~\added{\citet{Csay201472}} and other researchers using different AOPs~\added{\citep{Doll2004955,Sires2007373,Rosal20091061}}.

It should be pointed out that, for the longer treatment times in the case of plasma and for the higher adsorbed doses in the case of gamma radiation, HPLC analysis revealed that the above products in turn undergo degradation, as was expected, thus confirming their role as reaction intermediates along the oxidation chain of CFA. The progress of the oxidation reaction depends not only on the energy input but also on the organic compound initial concentration. Specifically we found that after 4 hours of plasma treatment or an absorbed dose of \SI{10}{kGy}, these intermediates were no longer detected in the experiments with an initial concentration of CFA equal to \SI{1e-4}{M}, while, under the same treatment conditions, they were still present when higher initial concentrations of CFA were used. We also detected and quantified the final product of oxidation, CO$_2$, by FT-IR analysis of the gas exiting the reactor, as described previously~\added{\citep{Marotta:PPAP201100036,Marotta:EPJAP2011}}. The results show that after \SI{300}{min} of treatment of a \SI{1e-3}{M} CFA solution in the NTP reactor, the amount of CO$_2$ produced corresponded only to \SI{5}{\percent} of the theoretical yield. The extent of mineralization, however, also depends drastically on the concentration of the organic pollutant being treated. Thus,~\added{\citet{Krishna201635}} found that treatment of the drug Verapamil in the same reactor as used in the present study afforded only \SI{14}{\percent} of mineralization in \SI{4}{h} when the drug initial concentration was \SI{5e-5}{M} but reached \SI{98}{\percent} in \SI{5}{h} when the initial concentration was \SI{1e-5}{M}. The CFA concentrations used in the present study are higher than those of Verapamil in the above mentioned experiments and, most importantly, are also much higher than found in CFA contaminated waters in the environment. Thus, both the degradation efficiency and the extent of mineralization are expected to be much higher for NTP treatment of CFA solutions at the environmental levels of pollution.

The results of this comprehensive analysis of complex reaction mixtures are summarized in Table~\ref{tbl:LCMS-P}, which reports MS and MS/MS data as well as the identity attributions for the various detected products. The products chemical structures are shown in Scheme~\ref{fig:Prod-G}.  

\pagebreak

\begin{table}[ht]
\centering
\begin{tabular}{p{5.7cm} p{1.5cm} p{1.6cm} p{1.2cm} p{1.2cm}} \hline\noalign{\smallskip}
\textbf{Metabolite} & \textbf{[M-H]}$^-$ (m/z) & \textbf{MS/MS} (m/z) & \textbf{Gamma} & \textbf{Plasma} \\ \hline\noalign{\smallskip}
\textbf{A}, \small{2-hydroxyisobutyric acid} & 103 &  & x & x \\
\textbf{B}, \small{2-(4-hydroxyphenoxy)-isobutyric acid} & 195 & 109, 85 & x & \\
\textbf{C}, \small{4-chlorocatechol} & 143 &  & x & \\
\textbf{D}, \small{4-chlorophenol} & 127 &  & x & x \\
\textbf{E}, \small{2-(4-chloro-hydroxyphenoxy)-isobutyric acid} & 229 & 143 & x & x \\
\textbf{F}, \small{2-chloro-5-hydroxycyclohexa-2,5-diene-1,4-dione} & 157 &  &  & x \\
\textbf{G}, \small{2-((4-chloro-3,6-dioxocyclohexa-1,4-dien-1-yl)oxy)-2-methylpropanoic acid} & 243 & 157 &  & x \\
\textbf{Z}, \small{unidentified} & 391 & 373, 259, 113 &  & x \\
\textbf{X}, \small{unidentified} & 235 & 103 & x &  \\[3pt] \hline
\end{tabular}
\caption{Results of LC-MS analysis of CFA solutions treated with gamma radiation and with a non-thermal plasma.}
\label{tbl:LCMS-P}
\end{table}

\section{Discussion of results}
\label{sec:D}

We start by pointing out the similarities and differences between the two technologies used in this work. In water radiolysis the energy transported by ionizing radiation ($\gamma$ photons, high energy electrons or $\alpha$ particles) excites and ionizes the water molecule leading to the production of $e_{aq}^-$, H$^\cdot$, $^\cdot$OH, HO$_2^\cdot$, OH$^-$, H$_3$O$^+$, H$_2$ and H$_2$O$_2$. In the case of $\gamma$ radiation and for $3 < \mathrm{pH} < 11$, the highest radiolytic yield are obtained for $^\cdot$OH, followed by $e_{aq}^-$~\added{\citep{Ershov08,Caer11}}.

The discharge in the NTP reactor is characterized by a rather homogeneous plasma envelope confined around the wires and not touching the liquid surface, with the sporadic occurrence of sparks. The current pulses observed in the positive corona phase are much higher than for negative corona~\added{\citep{Marotta:EPJAP2011}}. This is consistent with what is observed in dc discharges over a liquid, when $r/d << 1$ ($r$ is the electrode radius and $d$ the inter-electrode distance) where the discharge is a stable corona before electric breakdown. In the case of negative corona a glow-to-spark transition is observed, whereas for positive corona a streamer-to-spark discharge is observed~\added{\citep{Brugg09}}.

In a previous study~\added{\citep{Marotta:PPAP201100036}} we characterized and determined the reactive oxidizing species produced in the same NTP reactor and under the same experimental conditions used in the present work. Notably, it was concluded that $^\cdot$OH radicals and ozone are key species involved in the decomposition of the organic pollutant in this experimental set-up.
Specifically, we found that the rate of OH radical production, determined by monitoring the conversion of coumarin 3-carboxylic acid (CCA) into coumarin 7-hydroxy-3-carboxylic acid (CCA-7-OH)~\added{\citep{NEWTON2006473}} was \SI{4.4e-4}{\micro\mole/s}, whereas dissolved ozone in solution, determined spectrophotometrically by the indigotrisulfonate test~\added{\citep{BADER1981449}} was below the detection limit (\SI{1.4e-4}{mg/mL}).
Ozone concentration was also measured in the air above the aqueous phase. The data showed that consistently less ozone was present when the water contained the organic pollutant suggesting that in our apparatus ozone is used up by reaction with organic pollutant at the gas liquid interface.
Other important species found in air plasma-water interaction include HNO$_{2\,(aq)}$ and HNO$_{3\,(aq)}$ responsible for the acidification of the solution. UV radiation produced in the air is also present on the UVB (\SIrange{320}{400}{nm}) and UVC (\SIrange{100}{280}{nm}) wavelength ranges with typical intensities below \SI{50}{\micro\watt\per\square\centi\metre}~\added{\citep{0963-0252-25-5-053002}}.

\subsection{Kinetic model}\label{sec:4.1}

As observed in Figures~\ref{fig:F3}, the initial values of the ([CFA]/[CFA]$_0$) ratio follow an exponential law with time, with a degradation constant inversely proportional to the initial concentration. 
This dependency was explained~\added{\citep{Slater:JAP1981}} in terms of a mechanism of inhibition by products which compete for the reactive species involved in the process. This model, however, uses approximations and assumptions that are valid only in the initial steps of the degradation process.
Moreover, as the ratio decreases, the points deviate from the exponential fitting curve.

In order to explain the deviations observed at long reaction times and overcome the limitations of the previous model, we analysed this problem in the framework of an extended kinetic model. In this model a species, S, responsible for the degradation process (typically $\cdot$OH and H$_2$O$_2$) reacts with an organic compound, (HC)$_n$, and its consecutive intermediates (HC)$_{n-1}$, (HC)$_{n-2}$, \ldots, (HC)$_1$, until reaching a final stable product, C, (e.g. CO$_2$) through the reactions
\begin{equation}
 \mathrm{S} + \mathrm{(HC)}_j \rightarrow \mathrm{(HC)}_{j-1} + \text{other products} \qquad j=n,\ldots,1
\end{equation}
The species S is formed by either radiolysis or the interaction of the plasma with the liquid surface at a constant rate, R.\footnote{Note, however, that while for radiolysis we can assume a homogeneous rate of formation of S, in the case of the interaction of a plasma with a liquid surface this rate strongly depends on the distance from the surface. In this case a correct analysis requires a time and space dependent model and the consideration of diffusion processes. If we can neglect diffusion, however, and consider that R represents a space average rate, this discussion can be applied to a plasma-liquid interaction.}
In a closed system, the kinetic equations for these compounds are,

\begin{align}
\frac{d \left[\mathrm{(HC)_n}\right]}{dt} &= -k_n\left[\mathrm{(HC)_n}\right][\mathrm{S}] \label{eq:S1}\\
\frac{d \left[\mathrm{(HC)_{j-1}}\right]}{dt} &= 
\left\{k_j\left[\mathrm{(HC)_j}\right]-k_{j-1}\left[\mathrm{(HC)_{j-1}}\right]\right\}[\mathrm{S}]\qquad j = n,\ldots,2 \label{eq:S2}\\
\frac{d[C]}{dt} &= k_1\left[\mathrm{(HC)_1}\right][\mathrm{S}]  \label{eq:S3}
\end{align}
and
\begin{equation}
\frac{d[S]}{dt} = R - \sum_{j=1}^{n}k_j\left[\mathrm{(HC)_j}\right][\mathrm{S}]\,.  \label{eq:S4}
\end{equation}
Note that, while the concentrations $(\left[\mathrm{(HC)_j}\right],[C])$ change in time, we always have
\begin{equation}
\left[\mathrm{(HC)_n}\right]_0 \approx \sum_{j=1}^n \left[\mathrm{(HC)_j}\right](t) + [C](t)\,,  \label{eq:S5}
\end{equation}
where $\left[\mathrm{(HC)_n}\right]_0$ is the initial concentration. Using $k_m=\min(k_n,k_{n-1},\ldots,k_1)$ and $k_M=\max(k_n,k_{n-1},\ldots,k_1)$, the minimum and maximum rate coefficients respectively, and equation~(\ref{eq:S5}), we have the inequality
\begin{equation}
k_m \left\{\left[\mathrm{(HC)_n}\right]_0-[C](t)\right\} \le \sum_{j=1}^n k_j \left[\mathrm{(HC)_j}\right](t) \le 
k_M \left\{\left[\mathrm{(HC)_n}\right]_0-[C](t)\right\}\,,
\end{equation}
showing that the loss term in equation~\eqref{eq:S4} is always bound by values that do not depend on the concentrations of the intermediate products but only on the initial concentration, $\left[\mathrm{(HC)_n}\right]_0$, and the concentration of the final product, [C]. It can be shown that the last is a slowly growing function (in particular, much slower than the decomposition rate of (HC)$_n$) as it depends on the chain of intermediate products. As a next step, we define an average rate coefficient, $\bar{k}(t)$ as
\begin{equation}
\bar{k}(t) = \frac{\sum_{j=1}^n k_j\left[\mathrm{(HC)_j}\right](t)}{\sum_{j=1}^n \left[\mathrm{(HC)_j}\right](t)}\,, \label{eq:kbar}
\end{equation}
with $\bar{k}(0)=k_n$. This average rate coefficient depends on time through the concentrations of intermediate products but is bound by $k_m$ and $k_M$. By similar reasons as in the case of [C], it is also a slow function of time. 

Using equations~\eqref{eq:S5} and \eqref{eq:kbar} we write the loss term of equation~\eqref{eq:S4} as
\begin{equation}
\sum_{j=1}^{n}k_j\left[\mathrm{(HC)_j}\right][\mathrm{S}] =
\bar{k}(t)\left\{\left[\mathrm{(HC)_n}\right]_0-[C](t)\right\}[S]
\label{eq:S6}
\end{equation}
Substituting equation (\ref{eq:S6}) in (\ref{eq:S4}), the quasi-equilibrium value for S is
\begin{equation}
[S]_{\mathrm{qeq}}=\frac{R}{\bar{k}_t\left\{\left[\mathrm{(HC)_n}\right]_0-[C]_t\right\}}\,,\label{eq:Seq}
\end{equation}
where we have replaced the explicit dependency on time in $\bar{k}$ and $[C]$ by the index $t$ to indicate a slow variation. 
This value slowly increases with $[C]_t$. The concentration of S reaches this quasi-equilibrium value with a time constant of $\left(k_n\left[\mathrm{(HC)_n}\right]_0\right)^{-1}$. 

For $k_n\left[\mathrm{(HC)_n}\right]_0 t \gg 1$, we can replace the value of [S] in equations (\ref{eq:S1}-\ref{eq:S3}) by the quasi-equilibrium value, $[S]_\mathrm{qeq}$. If we neglect the slow time dependency of (\ref{eq:Seq}), the system (\ref{eq:S1}-\ref{eq:S3}) can be solved to obtain the approximate time-dependent solutions for (HC)$_n$ and sub-products. The solution for (HC)$_n$ is
\begin{equation}
\left[(HC)_n\right](t) \approx \left[(HC)_n\right]_0 
\exp\left(-\frac{k_n}{\bar{k}_t}\frac{R\,t}{\left[\mathrm{(HC)_n}\right]_0-[C]_t}\right). \label{eq:knew}
\end{equation}

Except for the $k_n/\bar{k}_t$ ratio and the presence of the concentration of the final product $[C]_t$, this expression is similar to the one proposed by~\added{\citet{Slater:JAP1981}} with a degradation constant inversely proportional to the initial concentration. The presence of $[C]_t$ and $\bar{k}_t$, however, shows that this value is not constant, depends directly on the concentration of the final product and, indirectly through $\bar{k}_t$, on the concentrations of the intermediates. Note that the presence of the ratio $k_n/\bar{k}_t$ means that the degradation of the organic compound, (HC)$_n$, depends not only on its own reaction rate, $k_n$, but on the ensemble of intermediates produced and the corresponding degradation rates, as can be seen by a direct substitution of equation~(\ref{eq:Seq}) in~(\ref{eq:S1}). This link between the reaction rates of all organic compounds happens because all of them compete for the same reactive species, $S$.

Note that the only approximations in obtaining equation~(\ref{eq:knew}) were to (i) consider a consecutive chain of intermediate products all reacting with S, (ii) replace [S] by $[S]_\mathrm{qeq}$ and, (iii) neglect its time dependency. The solutions of the remaining equations of the system (\ref{eq:S1}-\ref{eq:S3}) and a full discussion of this model will be published elsewhere. 

Although this model applies to a consecutive chain of similar reactions and the degradation of CFA proceeds through different parallel paths as indicated by the metabolites found and proposed in section~\ref{sec:DegMec}, it can be shown~\added{\citep{Pinhao17a}} that equation~(\ref{eq:Seq}) is still valid (with a different expression for $\bar{k}_t$) and the decomposition of CFA should follow equation~(\ref{eq:knew}).
The experimental results in Figure~\ref{fig:F3}, with an increase of the slope of the degradation curve for small values of the [CFA]/[CFA]$_0$ ratio (where we can expect a higher concentration of the final decomposition products), are in agreement with the present model. Note however that the $k_n/\bar{k}_t$ ratio also affects the slope. If this ratio is known or if it is possible to make an educated guess about this value, equation (\ref{eq:knew}) allows the concentration of the final product C to be obtained from an analysis of the slope of the experimental data.

In our case, without a previous knowledge of the kinetics and rate coefficients, we have assumed a constant value for $\bar{k}_t$, i.e. $k_n/\bar{k}_t=1$, and a linear growth of [C] with t, $[C]_t=m_C t$:
\begin{equation}
[\mathrm{CFA}](t) \approx [\mathrm{CFA}]_0 
\exp\left(-\frac{R\,t}{[\mathrm{CFA}]_0-m_C t}\right), \label{eq:kCFA}
\end{equation}
with $m_C$ a rate constant of formation.

The fitting of the experimental results to this equation is represented in Figure~\ref{fig:F3} by the solid lines. The corresponding fitting coefficients are compiled in Table~\ref{tbl:FitCte}:

\begin{table}[ht]
\centering
\begin{tabular}{cccc|ccc}
\hline
 & \multicolumn{3}{c}{\textbf{Gamma radiation}} & \multicolumn{3}{c}{\textbf{Plasma}} \\ 
\textbf{[CFA]$_0$} & $\mathbf{R}$ & $\mathbf{m_C}$ & \textbf{G$_{50}$} &
$\mathbf{R}$ & $\mathbf{m_C}$ & \textbf{G$_{50}$} \\
(M) & (\si{\micro M/min}) & (\si{\micro M/min}) & (\si{g/\kilo Wh}) &
(\si{\micro M/min}) & (\si{\micro M/min}) & (\si{g/\kilo Wh}) \\
\hline
\num{1e-4} & \num{5.7} & \num{3.1} & \num{15e1} & \num{2.7} & \num{0.28} & \num{1.2} \\
\num{5e-4} & \num{12} & \num{0.88} & \num{23e1} & \num{3.5} & \num{0.59} & \num{1.6} \\
\num{1e-3} & \num{12} & \num{0.52} & \num{22e1} & \num{2.9} & \num{0.74} & \num{1.4} \\
\hline
\end{tabular}
\caption{Fitting constants for equation~(\ref{eq:kCFA}) for the degradation of CFA by gamma radiation and plasma, and energy yield given by equation~(\ref{eq:eta}) for different values of initial concentration.}
\label{tbl:FitCte}
\end{table}

Except for gamma radiation with an initial concentration of \SI{1e-4}{M}, the values of the formation rate of the species responsible for the degradation are close to each other for each technology, as expected. This rate is much higher for the the gamma radiation installation than for the plasma reactor used. Note, however, that the rate of formation of the final product C is small and similar for both technologies. 
This outcome can be explained by a slow step in the sequence of reactions involving many intermediates along the oxidation of CFA to CO$_2$. The slow reacting intermediate might be the same or a different one in the two tested processes, activated by air plasma and by gamma-radiation, respectively.

\subsection{Decomposition products and degradation mechanisms}\label{sec:DegMec}

\added{\citet{Sires2007373}} studied the oxidation of clofibric acid using Fenton systems and proposed a reaction scheme in which CFA is first oxidized to 4-chlorophenol by rupture of the C(1)-O bond and then also yielding 2-hydroxyisobutyric acid. The hydroxyl attack on C(2) position leads to 4-chlorocatechol. \added{\citet{Doll2004955}} investigated the catalytic photodegradation products of CFA and proposed the same pathway together with the dechlorination reaction which supposedly yields 2-(4-hydroxyphenoxy)-isobutyric acid.

Based on these previous results and on the identification of the intermediate products described in section~\ref{sec:DecPr}, the possible degradation pathway and reaction intermediates for CFA oxidation in gamma irradiation and non-thermal plasma experiments are proposed in Scheme~\ref{fig:Prod-G}.
Note that in the case of gamma radiation all of the observed products (A, B, C, D and E) can be attributed to reactions with the hydroxyl radical whereas in the case of non-thermal plasma treatment ozone also appears to be involved in the oxidation process. Based on the results of a thorough study on the ozonation of phenolic compounds~\added{\citep{B301824P}} we propose that the product G, which is not observed in gamma radiolysis experiments, is due to a reaction with ozone. 

Interestingly, in the analysis of NTP treated CFA solutions, we never detected any product possibly deriving from reactions with reactive nitrogen species~\added{\citep{0963-0252-25-5-053002}}.


\renewcommand{\figurename}{Scheme}
\setcounter{figure}{0}

\begin{figure}[ht]
\centering\includegraphics[width=\textwidth]{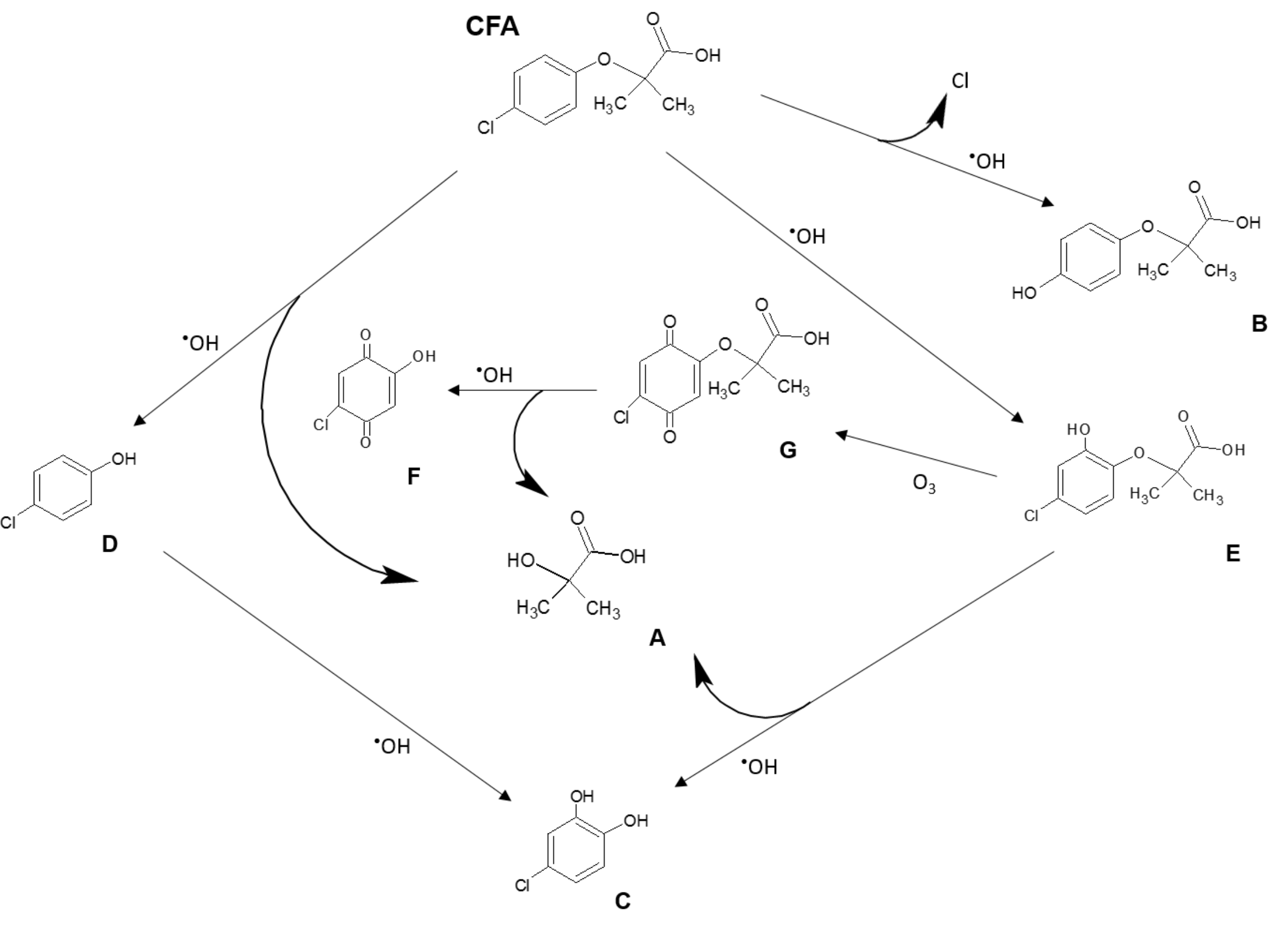}
\caption{Observed products and proposed mechanisms of degradation of CFA by gamma radiation and plasma treatment.}
\label{fig:Prod-G}
\end{figure}

\subsection{Energy yield}

In order to evaluate the energy yield of the two experimental systems for the degradation of clofibric acid, we started by analysing the power consumption in each system.
The power, $P$, consumed by the electrical discharge is estimated at \SI{1.6}{W}.  This power is used on the maintenance of the plasma and production of charges and reactive species that interact with the liquid. Considering the volume of aqueous solution treated this corresponds to a power per unit volume of \SI{22.9}{W/L}.  For gamma radiation, the dose rate used corresponds to a power deposited on the sample of \SI{0.5}{W/L} or \SI{1.5e-3}{W} on our samples. We can compute the energy yield, $G_{50}$, defined as the amount of CFA converted divided by the energy per unit volume required to reduce its concentration by 50\% (expressed in \si{g/kWh}) as~\added{\citep{Malik2010}},
\begin{equation}
G_{50} = \frac{1}{2}\frac{V}{P}\frac{c_0}{t_{50} } \,,
\label{eq:eta}
\end{equation}
where $V$ is the volume of the sample, $c_0$ the CFA initial concentration in \si{g/L} and $t_{50}$ is the time, expressed in hours, required to obtain \SI{50}{\%} degradation of CFA.

Using the fitting parameters reported in table~\ref{tbl:FitCte} we obtain the corresponding values for $G_{50}$.
In both systems the energy efficiency do not has a clear trend as a function of the initial concentration of CFA. 
The dimension of each independent set of data is not sufficient for a meaningful discussion of possible effects of [CFA]$_0$ on $G_{50}$. In the literature the dependence of $G_{50}$ on the initial concentration of the organic pollutant to be removed is not well characterized and understood. This is however not surprising considering equation~(\ref{eq:eta}): when $c_0$ increases, $t_{50}$ also increases, but the relationship between them depends on the treated pollutant.

The higher values of $G_{50}$ found in gamma radiation with respect to non-thermal plasma for the degradation of clofibric acid can be attributed to the higher penetration of gamma rays in the solution thus increasing the effective volume under treatment and the effective decomposition of the contaminant. The non-thermal plasma generates electrons and reactive species that penetrate the surface of the liquid, diffuse into the liquid phase and react with the organic compounds within the solution.

It is not possible, however, to extrapolate the conclusions reached in the comparison between the two experimental systems used in this study to a general comparison between the two technologies. Thus, the plasma reactor was specifically designed for achieving stable and reproducible conditions necessary to perform fundamental studies of plasma-liquid interactions and was not optimized with respect to energy efficiency. In addition, we did not consider, in the case of gamma radiation experiments, the electric power necessary to operate the gamma source.

\section{Conclusions}
\label{sec:C}

The main conclusions of this study on the degradation of CFA in aqueous solution, using gamma radiation and a non-thermal plasma, are:

\begin{itemize}
\item Both technologies are able to achieve high conversions of CFA;
\item In both cases, the degradation of CFA follows a quasi-exponential decay in time  which is well modelled by a kinetic scheme based on the competition of CFA and of all reaction intermediates for the reactive species generated in solution and that takes into account the amount of the end product formed;
\item Even though the conversion of CFA was high and CO$_2$ was detected among the products, mineralization was highly incomplete. A few degradation intermediates still having the aromatic ring with chlorine were detected and identified;
\item Some of the degradation products obtained using the two technologies, gamma radiation and air plasma, are different suggesting that different mechanisms are active in the two systems; and
\item In the systems studied the energy yield was found to be higher for the gamma radiation installation.
\end{itemize}

\section*{Acknowledgements}

This work was supported by COST Action TD1208, the Portuguese Fundação para a Ciência e a Tecnologia through projects RECI/AAG-TEC/0400/2012, \linebreak
UID/Multi/04349/2013 and UID/FIS/50010/2013 and the University of Padova through grant CPDA147395/14 (Progetto di Ricerca di Ateneo 2014).

\section*{References}

\bibliographystyle{elsarticle-harv}
\bibliography{CFA-plasma_gamma.bib}

\pagebreak
\section*{List of figure captions}

\begin{description}
\item[Figure 1:] Structure of the investigated compound, clofibric acid.
\item[Figure 2:] Chromatograms of solutions, with an initial clofibric acid concentration of \SI{1e-3}{M}, irradiated on a \added{$^{60}$Co} gamma source, as a function of the absorbed dose. The dose rate in the samples was \SI{1.8}{kGy/h}.
\item[Figure 3:] Degradation of clofibric acid at different initial concentrations using: (a) gamma irradiation and (b) non-thermal plasma treatment. Experimental values (points), exponential fit (dotted lines) and fit to the new kinetic model described in  section~\ref{sec:4.1} (solid lines).
\item[Figure 4:] Negative ion MS/MS spectra of intermediates (a) B , (b) E  and (c) G detected in the treatment of \SI{1e-3}{M} CFA with gamma radiation and/or with air non-thermal plasma (see Table~\ref{tbl:LCMS-P} and Scheme~\ref{fig:Prod-G} for details).
\item[Scheme 1:] Observed products and proposed mechanism of degradation of CFA by gamma radiation and plasma treatment.
\end{description}

\end{document}